\documentstyle[epsf]{mn}
\begin{document}

\title[Parameter extraction for a BSI $\Lambda$CDM model]
{Parameter extraction by {\it Planck} for a 
CDM model with broken scale invariance and cosmological constant}
\author[J.~Lesgourgues, S.~Prunet and D.~Polarski]
{Julien Lesgourgues,$^{1,4}$ Simon~Prunet$^{2}$ and David~Polarski$^{1,3}$\\
$^1$~{\it Laboratoire de Math\'ematiques et de Physique Th\'eorique, 
UPRES-A 6083 CNRS}\\
{\it Universit\'e de Tours, Parc de Grandmont, F-37200 Tours (France)}\\
$^2$~{\it Institut d'Astrophysique Spatiale,
b\^at. 121, 91405 Orsay cedex (France)}\\ 
$^3$~{\it D\'epartement d'Astrophysique Relativiste et de Cosmologie},\\
{\it Observatoire de Paris-Meudon, 92195 Meudon cedex (France)}\\
$^4$~{\it International School for Advanced Studies, SISSA-ISAS, 
Via Beirut 4, I-34014 Trieste (Italy)}\\}

\date{28 September 1998}


\maketitle

\begin{abstract}
We consider a class of spatially flat cold dark matter (CDM) models,
with a cosmological constant and a broken-scale-invariant (BSI)
steplike primordial spectrum of adiabatic perturbations, previously
found to be in very good agreement with observations. 
Performing a Fisher matrix analysis, we show that in case of a large
gravitational waves (GW) contribution some free parameters 
(defining the step) of our 
BSI model can be extracted with remarkable accuracy by the {\it Planck} 
satellite, thanks to the polarisation anisotropy measurements.
Further, cosmological parameters can still be found with very good 
precision, despite a larger number of free parameters than in the simplest 
inflationary models.
\end{abstract}


\section{Introduction}

Current observations of the large-scale structure in the Universe,
and of cosmic microwave background (CMB) anisotropies in particular, 
already allow a discrimination among different cosmological scenarios with 
increasing precision. Nevertheless, many possibilities
are still viable, with different assumptions concerning e.g. the matter 
content of the Universe, and the primordial (initial) fluctuation power 
spectrum. However, most scenarios should be excluded by cross-correlating
the forthcoming experiments, like, for instance, balloon and satellite 
measurements of small scale CMB anisotropies, and new reshift surveys
(Wang, Spergel \& Strauss 1999). The
most precise scheduled experiment for the measurement of the CMB anisotropies 
is the {\it Planck} satellite\footnote{For the instrumental 
specifications of the mission, see 
http://astro.estec.esa.nl/SA-general/Projects/Planck/}, the data from which
will very
likely favour a restricted family of cosmological scenarios, hopefully with a 
small number of free parameters. 

As the simplest CDM model with a flat primordial power spectrum
is already excluded, it is necessary to introduce 
some refinements either in the content of the Universe 
(i.e., in the transfer functions of matter and radiation), 
or in the generation of initial fluctuations (i.e., in the case
of inflationary models, in the primordial power spectrum). 
By now, the simplest variant favoured by experimental data seems to be 
that of a flat universe with a cosmological
constant, $\Omega_m+\Omega_\Lambda=1$, and a scale-invariant primordial 
(or slightly 
tilted) power spectrum (Kofman, Gnedin \& Bahcall 1993; Bagla, Padmanabhan
\& Narlikar 1996; Ostriker \& Steinhardt 1995; Lineweaver 1998). 
In two recent papers (Lesgourgues, Polarski \& Starobinsky 1998a, 1998b,
further referred as LPS1, LPS2), 
the combination of this $\Lambda$CDM scenario with an inflationary 
model introduced by Starobinsky (1992), predicting a 
broken-scale-invariant (BSI) steplike primordial power spectrum, was 
investigated. 
In LPS1, the case of adiabatic primordial fluctuations was considered when 
the contribution of the tensorial fluctuations to the CMB 
anisotropies is negligible. In LPS2, the
possible contribution of gravitational waves to the CMB anisotropy was 
taken into account: it is a most interesting peculiarity of these models 
that these distinct cases are possible and were shown to be viable regarding 
observations. 
Using polarization measurements on the precision level scheduled for 
{\it Planck}, this large GW contribution will allow accurate parameter 
extraction.
Briefly, the motivations for considering steplike models are the following.
First, an even better agreement with the data can be found
than in the case of a flat or tilted spectrum,
inside a wider region of the cosmological parameter space.
Second, a few authors point out the possible observational evidence for a 
spike in the matter power spectrum at
$k\simeq 0.05~h~{\rm Mpc}^{-1}$ (Einasto et al. 1997a, 1997b, 1997c; 
Retzlaff et al. 1997; Gazta\~naga \& Baugh 1998). This is still a point of 
debate; however, as our BSI model predicts a similar feature, it would clearly 
be an excellent remaining candidate if the spike were 
to be confirmed by future 
redshift surveys. 

One could argue that BSI inflationary models, by introducing additional
free parameters in the primordial power spectrum, just increase the 
degeneracy among different scenarios; that instead of making real predictions,
like the simplest slow-roll models, they are just introduced {\it ad hoc},
in order to fit any observations; and finally, that in the case of BSI 
primordial spectra, the prospect of extracting the cosmological 
parameters at the per cent level with {\it Planck} would collapse.
However, we recall that our model is based on simple assumptions concerning 
the inflaton potential, and cannot be tailored at will in order to fit any 
given observational data. On the contrary, it predicts a very peculiar 
observable feature in the matter power spectrum at intermediate scales 
($\simeq 125~h^{-1}~{\rm Mpc}$) while it makes, of course, predictions on all 
scales both for the matter and the radiation power spectrum. 
Even when only the radiation power spectrum is considered, we
show in this work the following points: 
\begin{itemize}
\item the future {\it Planck} results should easily discriminate
between our BSI model and other scale-free models.
\item assuming that this model is indeed realized in nature,
in spite of four additional degrees of freedom in the theory compared 
to the simplest versions of inflation, {\it Planck} should still be able 
to measure accurately both the cosmological and the inflationary parameters. 
Furthermore, it turns out that one of the inflationary parameters, $p$, which 
defines both the height and the shape of the step, should be constrained 
with remarkably high precision, a fact which could be of significant interest 
for building particle-physics-motivated inflationary models.
\end{itemize}

\section{The model}

We assume for simplicity that our Universe is known to be spatially
flat, and that neutrino mass and reionisation can both be neglected
(relaxing these assumptions would of course increase the uncertainties
on all parameters). Then,
our model contains three cosmological parameters 
($h$, $\Omega_{\Lambda}$, $\Omega_b$), 
and five inflationary parameters, which can be understood as follows:
\begin{itemize}
\item
the power spectrum of adiabatic perturbations 
has a scale-invariant tilt $n_s$ on large scales, $k<k_0$, undergoes a break 
(the shape of which is defined by one single parameter $p$) at $k \geq k_0$, 
and is finally flat on small scale $k \gg k_0$. 
The ratio between the power spectrum on the small-scale plateau and at $k_0$\ is given by $p^{-2}$. 
\item
the spectrum of GW has no break at $k_0$, while the 
tensor tilt on small scale $k>k_0$\
is irrelevant for our purpose, because the corresponding
contribution to the $C_l$'s is negligible. 
Using the slow-roll conditions valid on large scales, the scale-dependent 
tilt $n_T(k)$ for $k\leq k_0$ can be found as a 
function of $n_S$ and $n_T(k_0)$.
\item
as the slow-roll approximation is still valid for large-scale 
perturbations,
at $k=k_0$ one can relate the amplitude of the GW power spectrum to the 
dimensionless parameter $H_{k_0}^2 G$, and the scalar power spectrum amplitude 
to $H_{k_0}^2 G / n_T(k_0)$.
\end{itemize}
In summary, the five free inflationary parameters are:
\begin{description}
\item[1.]
$H_{k_0}^2 G$, the overall dimensionless normalization factor. Varying
$H_{k_0}^2 G$\ (all other parameters being fixed)
is exactly similar to varying the commonly used $Q_{10}$, the 10-th
multipole of the temperature anisotropy power spectrum (Lineweaver \&
Barbosa 1998).
Hence, we will further use this parameter instead of $H_{k_0}^2 G$.
\item[2.]
$k_0$, the scale of the break.
\item[3.]
$p$, which defines the break's amplitude and shape.
\item[4.] 
$n_S$, the scale-invariant scalar tilt on scales $k<k_0$.
\item[5.]
$n_T(k_0)$, the (effective) tensor tilt at $k_0$. 
\end{description}
The usual tensor-to-scalar ratio $C^T_{10}/C^S_{10}$\ does not appear
in a natural way in this description. 
For fixed values of the cosmological parameters, there is a  
non-trivial dependence of $C^T_{10}/C^S_{10}$\ on $n_T(k_0)$\ and $n_S$. 
Therefore, fixing the parameters $n_T(k_0)$ and $n_S$ fixes the ratio 
$C^T_{10}/C^S_{10}$ as well.

In previous studies (LPS1, LPS2), $\Omega_b h^2=0.015$ was assumed while 
$k_0$\ was fixed by the Einasto et al. cluster data (the spike in the matter 
power spectrum at $k=0.05~h~$Mpc$^{-1}$ requires $k_0=0.016~h~$Mpc$^{-1}$).
Two possibilities for the scalar tilt were investigated:
\begin{description}
\item[A.] $n_s\approx 1$, which implies $d n_T/d \ln k \approx n_T^2$.
\item[B.] $n_s\approx 1+n_T={\rm constant}$, which implies $n_T(k)= 
n_T(k_0)= {\rm constant}$.
\end{description}
Further, a double normalization was performed to both 
$Q_{10}=18~\mu K$\ (Bennett et al. 1996) 
and $\sigma_8=0.60~\Omega^{-0.56}$\ 
(White, Efstathiou \& Frenk 1993). 
With these constraints, the remaining free parameter space
was three-dimensional:
($h$, $\Omega_{\Lambda}$, $n_T(k_0)$), or equivalently 
($h$, $\Omega_{\Lambda}$,$C^T_{10}/C^S_{10}$). 
In both cases A and B, the preferred regions following from the current 
observations were found.
Now, we choose one point inside each allowed region, corresponding
to $(h, \Omega_{\Lambda}, C^T_{10}/C^S_{10})=(0.7,0.7,0.8)$\ for both cases 
A and B. Assuming that each of these two points describes the 
``true'' cosmological scenario, we perform a Fisher matrix analysis with 
eight free parameters ($h$, $\Omega_{\Lambda}$, $\Omega_b$, $Q_{10}$, $k_0$, 
$p$, $n_S$, $n_T(k_0)$).

\section{The Fisher matrix}

Using the CMB Boltzmann code {\sc CMBFAST} (Seljak \& Zaldarriaga 1996), 
we compute the derivative of the $C_l$'s with respect to each parameter 
$\theta_{i,~i=1,.,8}$. 
The Fisher matrix (Jungman et al. 1996a, 1996b; Tegmark, Taylor \&
Heavens 1997; see also Bond, Efstathiou \& Tegmark 1997;
Copeland, Grivell \& Liddle 1998; Stompor \& Eftathiou 1998;
Eisenstein, Hu \& Tegmark 1998) is then obtained by
adding the derivatives, weighted by the inverse of the 
covariance matrix of the estimators of the polarized and unpolarized
CMB power spectra for the {\it Planck} satellite mission ,
$\rm{Cov}(C_{\ell}^X,C_{\ell}^Y)$: 
\begin{equation}
F_{ij}=\sum_{\ell=2}^{+\infty}\sum_{X,Y}
\frac{\partial C_{\ell}^X}{\partial \theta_i}
\rm{Cov}^{-1}\left(C_{\ell}^X,C_{\ell}^Y\right)
\frac{\partial C_{\ell}^Y}{\partial \theta_j}~,
\end{equation}
where $\{X,Y\} \in \{T,E,TE\}$\ (Kamionkowski, Kosowsky, \& Stebbins
1997, Zaldarriaga, Spergel \& Seljak 1997;
Prunet, Sethi \& Bouchet, 1998a, 1998b). 
The meaning of $F_{ij}$\ is the following.
Assuming that a fit to the {\it Planck} data yields a maximum likelihood
for the model under consideration (for which the derivatives were computed),
the 1-$\sigma$\ confidence region in the eight-dimensional parameter space
would be inside the ellipsoid (Press et al. 1989): 
\begin{equation}
\sum_{i,j} \Delta \theta_i  \Delta \theta_j F_{ij}=9.3~.
\end{equation}
Using $F_{ij}$ (or the dimensionless Fisher matrix
$\tilde{F}_{ij} \equiv \theta_i \theta_j F_{ij}$), 
one can also compute the allowed region in lower dimensional 
cuts of the parameter space, making no assumptions on  
other parameters. In particular, the 1-$\sigma$\ uncertainty on a single 
parameter is just the square root of a diagonal coefficient
of the inverse Fisher matrix:
\begin{equation}
\Delta \theta_i = \sqrt{(F^{-1})_{ii}}~.
\end{equation}

Each multipole will be measured by Planck with a precision of the
order of 1\%. As there are many more independent measurements
than free parameters, one naively
expects the parameter extraction to be much more precise. 
However, in general, the parameters are degenerate, i.e.,
some combinations of parameters produce a very weak change in the
$C_l$ curve. Hence, even when some other combinations can be measured with 
very high precision, each parameter separately is constrained only at the
percent level (unless its effect is ``orthogonal'' to the other ones).

A usefull way to express the results of the Fisher matrix analysis, 
which does not depend on a particular choice of basis in the parameter space, 
and contains the most refined constraints that can be deduced from the 
experiment, is to diagonalize $\tilde{F}_{ij}$. The eigenvectors correspond to
the axes of the likelihood ellipsoid, 
and the inverse square root of the eigenvalues to the 1-$\sigma$ 
relative uncertainties on each eigenvector. Eigenvectors with the
smallest uncertainties are the best constrained parameter combinations 
(they generate maximal changes in the anisotropy curve).
Eigenvectors with the largest uncertainties are the worst-constrained 
combinations (they generate minimal changes), and are generally 
called degenerate directions in parameter space.

In computing the covariance matrix of the CMB power spectra, we accounted
for the presence of foregrounds (both polarized and unpolarized) in
the measurement of the CMB power spectra, using the method described
in Bouchet, Prunet \& Sethi 1998 (see also Prunet, Sethi \& Bouchet 
1998a, 1998b). The derivatives of the spectra have been computed with an
error-minimizing routine derived from {\it Numerical Recipies} 
(Press et al. 1989). 

\section{Results}

The uncertainty on each parameter is presented in Table 1.
A and B stand for the two models previously mentioned. T is a tilted
model, with only three inflationary parameters instead
of five, namely ($Q_{10}$, $n_S$, $n_T(k_0)$), with the same values as in 
model B. 
Of course, one should keep in mind that all the uncertainties quoted
in Table 1 would increase if the space of free cosmological parameters 
were enlarged. In the lines without a $\times$-sign and with italic numbers, 
the polarisation measurement is not taken into account. 
The main conclusions to be drawn from the table are the following.

\begin{table*} \label{table1}
\begin{minipage}{120mm}

\caption{In the upper part of the table, we give the parameter values
for the chosen models: two BSI models A and B with 8 free parameters, 
and one tilted model T, with 6 free parameters. We also indicate
the related value of $C_{10}^T/C_{10}^S$.
The corresponding
relative 1-$\sigma$\ uncertainties, $\Delta \theta_i / \theta_i$, are given
in the lower part, in percent. 
In the lines without a $\times$-sign and with italic numbers, 
the polarisation measurement
is not taken into account. The uncertainty on $C_{10}^T/C_{10}^S$
was not calculated, but it is of the same order as the one on $n_T(k_0)$ 
since in a first order description, $C^T_{10}/C^S_{10}$ is 
approximately proportional to $n_T(k_0)$.}

\begin{tabular}{||l||c|c|c|c|c|c|c|c|c|c||}
\hline \hline
&\multicolumn{3}{c|}{cosmological par.}
&&\multicolumn{5}{c|}{inflationary parameters} 
& related \\ 
\cline{2-4} \cline{6-10}
model & $h$ & $\Omega_\Lambda$ & $\Omega_b$ &&  $Q_{10}$ & $k_0$ 
& $p$ & $n_s$ & $n_T(k_0)$ & $C_{10}^T/C_{10}^S$\\
\hline 
\hline 
A       & 0.7 & 0.7 & 0.03 && 18$~\mu K$& 0.016$~h$Mpc$^{-1}$
& 0.615 & 1 & -0.12 & 0.8 \\
\hline
B       & 0.7 & 0.7 & 0.03 && 18$~\mu K$ & 0.016$~h$Mpc$^{-1}$
& 0.51 & 0.825 & -0.175 & 0.8 \\
\hline
T  & 0.7 & 0.65& 0.03 && 18$~\mu K$& / & / & 0.825 & -0.175 & 0.8 \\
\hline
&\multicolumn{10}{c||}{relative 1-$\sigma$\ uncertainty (\%)} \\
\hline
A $\times$ &0.72 &0.94 &0.86 &&3.2 &0.82 &0.097 &6.0 &6.3 & \\ \hline
A          &\it 0.92&\it 1.2&\it 1.1&&\it 4.5&\it 1.1&\it 0.57&\it 15&\it 18& \\ \hline
B $\times$ &0.65 &0.85 &0.79 &&3.2 &0.75 &0.088 &9.3 &6.0 & \\ \hline
B          &\it 0.78&\it 1.0&\it 0.90&&\it 3.7&\it 0.90&\it 0.80&\it 44&\it 29& \\ \hline
T $\times$ &0.72 &0.93 &1.0  &&0.19&/    &/     &0.29&0.60& \\ \hline
T          &\it 0.86&\it 1.11&\it 1.21&&\it 0.24&/&/&\it 0.34&\it 0.70& \\ \hline
\hline

\end{tabular}
\end{minipage}
\end{table*}

 First, the three cosmological parameters are constrained with almost
the same precision for the tilted and for the BSI models; this means that
the step parameters ($k_0$, $p$) do not ``conspire'' with the parameters ($h$, 
$\Omega_\Lambda$, $\Omega_b$) in order to create directions of degeneracy.
Hence, in general,
the one percent parameter extraction proposed by {\it Planck} is
not affected in the case of BSI steplike models.

The situation is somewhat different for the inflationary parameters.
The normalization and tilts, ($Q_{10}$, $n_S$, $n_T$), 
appear less constrained; on the other hand, the step parameters, 
($k_0$, $p$), can be predicted with excellent accuracy, up to 
a 0.09 \% 1-$\sigma$ errorbar for $p$ ! These results can be easily understood,
especially if one keeps in mind that the best constraints come from 
high $l$ multipoles, for which the cosmic variance can be
neglected. For tilted models, the scalar tilt enters in all
multipoles, and can be accurately determined from high $l$'s; 
the two remaining inflationary parameters ($Q_{10}$, $n_T$) 
have a similar effect on high $l$'s (since $n_T$\ is proportional to the
tensor to scalar ratio), but measurements of the $C_l$'s for small $l$'s and 
polarisation measurements reduce the degeneracy. 
For our BSI models, the scalar tilt cannot be deduced from high $l$'s (it is 
defined at $k<k_0$, i.e., mainly $l<100$); the three
parameters ($Q_{10}$, $n_S$, $n_T(k_0)$) combine into several
degeneracy directions that can be resolved only by small $l$ measurements, 
so the precision remains poor. The situation is exactly opposite for the step
parameters ($k_0$, $p$), which have the crucial property of playing
a role {\it only} at $l>150$. Hence, they are only marginally affected by
cosmic variance. Further, $p$ is orthogonal to the degeneracy directions, 
and can be extracted with great precision.

All these features can be deduced with more accuracy from the Fisher matrix
diagonalization, given in Table 2.
\begin{table*} \label{table2}
\begin{minipage}{120mm}
\caption{Orthonormal eigenvectors of the dimensionless Fisher matrix 
$\tilde{F}_{ij}$, with their 1-$\sigma$\ uncertainty (in percent).
The first lines show some combinations of the parameters 
that can be recovered with a precision
much smaller than 1 \%. 
The last lines correspond to the directions of degeneracy in parameter space.}
\begin{tabular}{||l|l|l|l|l|l|l|l||c||}
\hline
\hline
\multicolumn{8}{||c||}{eigenvector}  & uncertainty  \\
\cline{1-8}
  $\frac{\Delta h}{h}$ 
& $\frac{\Delta \Omega_\Lambda}{\Omega_\Lambda}$ 
& $\frac{\Delta \Omega_b}{\Omega_b}$ 
& $\frac{\Delta Q_{10}}{Q_{10}}$ 
& $\frac{\Delta k_0}{k_0}$ 
& $\frac{\Delta p}{p}$ 
& $\frac{\Delta n_s}{n_s}$ 
& $\frac{\Delta n_T(k_0)}{n_T(k_0)}$   
& (\%) \\
\hline
\hline
-0.4 & 0.3 & 0.1 & 0.3 &-0.1 &-0.7 & 0.2 &-0.2 & 0.03\\
\hline
-0.6 & 0.6 & -   &-0.1 & 0.2 &-0.4 & -   & 0.1 & 0.05\\
\hline
-0.5 &-0.2 & 0.1 & -   &-0.8 & 0.2 & -   & -   & 0.1 \\
\hline
0.2  &  -  & -   & 0.6 & 0.1 & 0.4 & 0.3 &-0.5 & 0.2 \\
\hline
0.3  & 0.3 & 0.8 &-0.1 &-0.2 & -   & -   & -   & 0.3 \\
\hline
-0.4 &-0.6 & 0.5 & -   & 0.5 & -   & -   & -   & 1.5 \\
\hline
-    &  -  & -   & 0.7 & -   & -   &-0.2 & 0.7 & 4.5 \\
\hline
-    &  -  & -   &-0.1 & -   & -   & 0.8 & 0.5 & 11  \\
\hline
\hline
\end{tabular}
\end{minipage}
\end{table*}   
The first lines give parameter combinations that are constrained with
great precision; the last lines indicate the directions of degeneracy in
parameter space. It is straightforward to see that the inflationary parameter 
$p$ contributes only to the first four lines. 
Therefore, it doesn't suffer from any degeneracy, and is the
best constrained parameter. 
The 5th and 6th lines show that a change in $k_0$\ (resulting in a slight 
change in the location of the first acoustic peak, through the change in the 
primordial power spectrum), can be cancelled by a change in the
cosmological parameters (i.e., in the sound horizon scale).
The 6th eigenvector has a 1.5\% uncertainty: this is already a small
degeneracy, and $h$, $\Omega_{\Lambda}$, $\Omega_b$, $k_0$ 
are not as well constrained as $p$.
The last two lines show the large degeneracy between 
$Q_{10}$, $n_S$\ and $n_T(k_0)$.
 
Let us compare the uncertainties when polarisation measurements are taken 
into account and when they are not. 
Usually, the addition of polarized spectra leads to a small precision 
increase (by a factor $<2$) for all parameters 
which were not part of a specific degeneracy, as can be seen e.g. 
for tilted models, by comparing the (T) and (T$\times$) data in Table 1. In 
our BSI model, 
the precision on ($n_S$, $n_T(k_0)$, $p$) increases by a much larger factor,
and even by one order of magnitude for the parameter $p$ ! So, measuring 
the polarisation 
is even more important when one considers primordial spectra with
additional free parameters
(i.e., additional potential degeneracies to remove). One could be
surprised by the factor 10 found for $p$ in model B. 
In fact, when polarisation is not taken into account, 
$p$ enters into a single combination of parameters leading to a degeneracy. 
When polarisation is added, this degeneracy is supressed and, as we saw, 
$p$\ doesn't enter into any degeneracy at all. This mechanism
is illustrated in figure 1.

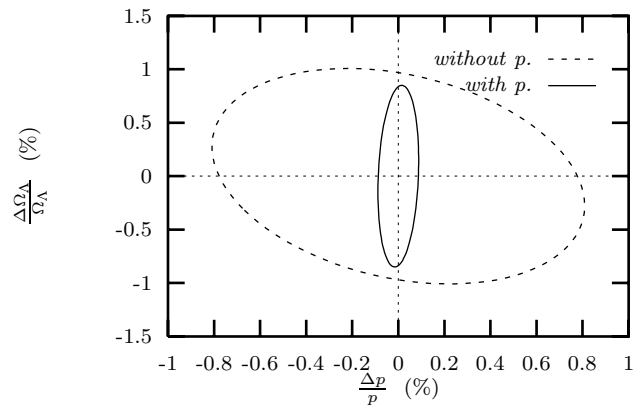
\begin{figure} \label{figure1}
\caption{1-$\sigma$ likelihood regions in the 
($\Delta p/p$, $\Delta \Omega_\Lambda/\Omega_\Lambda$) plane, with and without 
including polarisation measurement.
The only degeneracy involving $p$ is removed by the introduction of
polarization measurement. Therefore, the ellipse appears vertical in all 
($\Delta p/p$, $\Delta \theta_i/\theta_i$) plots.}
\setlength{\unitlength}{0.1bp}
\special{!
/gnudict 40 dict def
gnudict begin
/Color false def
/Solid false def
/gnulinewidth 5.000 def
/vshift -33 def
/dl {10 mul} def
/hpt 31.5 def
/vpt 31.5 def
/M {moveto} bind def
/L {lineto} bind def
/R {rmoveto} bind def
/V {rlineto} bind def
/vpt2 vpt 2 mul def
/hpt2 hpt 2 mul def
/Lshow { currentpoint stroke M
  0 vshift R show } def
/Rshow { currentpoint stroke M
  dup stringwidth pop neg vshift R show } def
/Cshow { currentpoint stroke M
  dup stringwidth pop -2 div vshift R show } def
/DL { Color {setrgbcolor Solid {pop []} if 0 setdash }
 {pop pop pop Solid {pop []} if 0 setdash} ifelse } def
/BL { stroke gnulinewidth 2 mul setlinewidth } def
/AL { stroke gnulinewidth 2 div setlinewidth } def
/PL { stroke gnulinewidth setlinewidth } def
/LTb { BL [] 0 0 0 DL } def
/LTa { AL [1 dl 2 dl] 0 setdash 0 0 0 setrgbcolor } def
/LT0 { PL [] 0 1 0 DL } def
/LT1 { PL [4 dl 2 dl] 0 0 1 DL } def
/LT2 { PL [2 dl 3 dl] 1 0 0 DL } def
/LT3 { PL [1 dl 1.5 dl] 1 0 1 DL } def
/LT4 { PL [5 dl 2 dl 1 dl 2 dl] 0 1 1 DL } def
/LT5 { PL [4 dl 3 dl 1 dl 3 dl] 1 1 0 DL } def
/LT6 { PL [2 dl 2 dl 2 dl 4 dl] 0 0 0 DL } def
/LT7 { PL [2 dl 2 dl 2 dl 2 dl 2 dl 4 dl] 1 0.3 0 DL } def
/LT8 { PL [2 dl 2 dl 2 dl 2 dl 2 dl 2 dl 2 dl 4 dl] 0.5 0.5 0.5 DL } def
/P { stroke [] 0 setdash
  currentlinewidth 2 div sub M
  0 currentlinewidth V stroke } def
/D { stroke [] 0 setdash 2 copy vpt add M
  hpt neg vpt neg V hpt vpt neg V
  hpt vpt V hpt neg vpt V closepath stroke
  P } def
/A { stroke [] 0 setdash vpt sub M 0 vpt2 V
  currentpoint stroke M
  hpt neg vpt neg R hpt2 0 V stroke
  } def
/B { stroke [] 0 setdash 2 copy exch hpt sub exch vpt add M
  0 vpt2 neg V hpt2 0 V 0 vpt2 V
  hpt2 neg 0 V closepath stroke
  P } def
/C { stroke [] 0 setdash exch hpt sub exch vpt add M
  hpt2 vpt2 neg V currentpoint stroke M
  hpt2 neg 0 R hpt2 vpt2 V stroke } def
/T { stroke [] 0 setdash 2 copy vpt 1.12 mul add M
  hpt neg vpt -1.62 mul V
  hpt 2 mul 0 V
  hpt neg vpt 1.62 mul V closepath stroke
  P  } def
/S { 2 copy A C} def
end
}
\begin{picture}(2519,1511)(0,0)
\special{"
gnudict begin
gsave
50 50 translate
0.100 0.100 scale
0 setgray
/Helvetica findfont 100 scalefont setfont
newpath
-500.000000 -500.000000 translate
LTa
600 856 M
1736 0 V
1468 251 M
0 1209 V
LTb
600 251 M
63 0 V
1673 0 R
-63 0 V
600 453 M
63 0 V
1673 0 R
-63 0 V
600 654 M
63 0 V
1673 0 R
-63 0 V
600 856 M
63 0 V
1673 0 R
-63 0 V
600 1057 M
63 0 V
1673 0 R
-63 0 V
600 1259 M
63 0 V
1673 0 R
-63 0 V
600 1460 M
63 0 V
1673 0 R
-63 0 V
600 251 M
0 63 V
0 1146 R
0 -63 V
774 251 M
0 63 V
0 1146 R
0 -63 V
947 251 M
0 63 V
0 1146 R
0 -63 V
1121 251 M
0 63 V
0 1146 R
0 -63 V
1294 251 M
0 63 V
0 1146 R
0 -63 V
1468 251 M
0 63 V
0 1146 R
0 -63 V
1642 251 M
0 63 V
0 1146 R
0 -63 V
1815 251 M
0 63 V
0 1146 R
0 -63 V
1989 251 M
0 63 V
0 1146 R
0 -63 V
2162 251 M
0 63 V
0 1146 R
0 -63 V
2336 251 M
0 63 V
0 1146 R
0 -63 V
600 251 M
1736 0 V
0 1209 V
-1736 0 V
600 251 L
LT0
LT2
2033 1297 M
180 0 V
888 727 M
27 -24 V
28 -24 V
31 -23 V
33 -22 V
34 -21 V
36 -20 V
38 -19 V
39 -18 V
41 -17 V
41 -16 V
43 -14 V
43 -13 V
44 -11 V
44 -10 V
45 -8 V
44 -7 V
45 -5 V
44 -4 V
44 -1 V
43 -1 V
42 2 V
41 3 V
40 4 V
39 6 V
37 8 V
35 9 V
34 11 V
32 13 V
30 13 V
27 15 V
26 17 V
23 17 V
20 19 V
18 20 V
16 21 V
12 21 V
10 23 V
7 23 V
5 24 V
1 25 V
-1 25 V
-4 25 V
-7 26 V
-10 25 V
-12 26 V
-15 26 V
-18 26 V
-20 25 V
-23 25 V
-25 24 V
-28 24 V
-29 23 V
-32 23 V
-34 21 V
-35 21 V
-37 20 V
-39 18 V
-39 18 V
-42 16 V
-42 15 V
-43 13 V
-43 12 V
-44 11 V
-45 9 V
-44 8 V
-45 6 V
-44 4 V
-44 3 V
-43 1 V
-43 -1 V
-42 -2 V
-40 -4 V
-39 -5 V
-38 -7 V
-37 -9 V
-34 -10 V
-33 -11 V
-31 -13 V
-29 -15 V
-26 -15 V
-25 -17 V
-21 -18 V
-20 -20 V
-16 -20 V
-14 -21 V
-12 -23 V
-8 -23 V
-6 -23 V
-3 -24 V
0 -25 V
3 -25 V
5 -26 V
9 -26 V
11 -25 V
13 -26 V
17 -26 V
19 -25 V
21 -25 V
24 -25 V
LT0
2033 1197 M
180 0 V
1455 513 M
5 1 V
4 2 V
5 3 V
5 5 V
5 6 V
5 7 V
4 9 V
5 10 V
5 11 V
4 12 V
4 14 V
4 14 V
4 16 V
4 16 V
4 17 V
3 18 V
3 19 V
3 20 V
3 20 V
2 20 V
2 21 V
2 21 V
2 22 V
1 22 V
1 21 V
0 22 V
1 22 V
0 21 V
-1 21 V
0 21 V
-1 20 V
-1 20 V
-2 19 V
-2 19 V
-2 17 V
-2 17 V
-3 16 V
-3 15 V
-3 14 V
-3 13 V
-4 12 V
-4 11 V
-4 9 V
-4 8 V
-4 7 V
-4 5 V
-5 5 V
-5 2 V
-4 2 V
-5 0 V
-5 -1 V
-5 -3 V
-5 -4 V
-5 -6 V
-4 -6 V
-5 -8 V
-5 -10 V
-4 -10 V
-5 -12 V
-4 -13 V
-4 -14 V
-4 -15 V
-4 -16 V
-4 -17 V
-4 -17 V
-3 -19 V
-3 -19 V
-3 -20 V
-2 -20 V
-2 -21 V
-2 -21 V
-2 -21 V
-1 -22 V
-1 -21 V
-1 -22 V
-1 -22 V
0 -21 V
0 -22 V
1 -21 V
1 -20 V
1 -20 V
1 -20 V
2 -19 V
2 -18 V
2 -17 V
3 -16 V
2 -16 V
3 -14 V
4 -14 V
3 -12 V
4 -11 V
4 -10 V
4 -9 V
4 -8 V
4 -6 V
5 -5 V
4 -3 V
5 -2 V
5 -1 V
stroke
grestore
end
showpage
}
\put(1973,1197){\makebox(0,0)[r]{{\it with p.}}}
\put(1973,1297){\makebox(0,0)[r]{{\it without p.}}}
\put(1468,51){\makebox(0,0){$\frac{\Delta p}{p}~~(\%)$}}
\put(100,855){%
\special{ps: gsave currentpoint currentpoint translate
270 rotate neg exch neg exch translate}%
\makebox(0,0)[b]{\shortstack{$\frac{\Delta \Omega_\Lambda}{\Omega_\Lambda}~~(\%)$}}%
\special{ps: currentpoint grestore moveto}%
}
\put(2336,151){\makebox(0,0){1}}
\put(2162,151){\makebox(0,0){0.8}}
\put(1989,151){\makebox(0,0){0.6}}
\put(1815,151){\makebox(0,0){0.4}}
\put(1642,151){\makebox(0,0){0.2}}
\put(1468,151){\makebox(0,0){0}}
\put(1294,151){\makebox(0,0){-0.2}}
\put(1121,151){\makebox(0,0){-0.4}}
\put(947,151){\makebox(0,0){-0.6}}
\put(774,151){\makebox(0,0){-0.8}}
\put(600,151){\makebox(0,0){-1}}
\put(540,1460){\makebox(0,0)[r]{1.5}}
\put(540,1259){\makebox(0,0)[r]{1}}
\put(540,1057){\makebox(0,0)[r]{0.5}}
\put(540,856){\makebox(0,0)[r]{0}}
\put(540,654){\makebox(0,0)[r]{-0.5}}
\put(540,453){\makebox(0,0)[r]{-1}}
\put(540,251){\makebox(0,0)[r]{-1.5}}
\end{picture}
\end{figure}

\section{Conclusion}
In this paper, we considered an inflationary model with BSI 
primordial spectrum and we investigated the precision with which the 
cosmological parameters and the free inflationary parameters could be 
extracted by the {\it Planck} satellite.
We first conclude that in the framework of the BSI steplike models considered 
here, the extraction of cosmological parameters can be as precise as 
in the case of tilted models.
The step parameters $p$ and $k_0$ can be constrained with excellent
accuracy, especially $p$, the effect of which on the $C_l$'s can be easily 
distinguished from the effect of any parameter combinations. 
There is no degeneracy with tilted models, which are special cases of 
our model with respect to the CMB anisotropies whenever 
$k_0\geq 0.25~h~{\rm Mpc}^{-1}$.
Further, if this class of models (or 
some other BSI model) were ever confirmed by future observations, it would be
reasonnable to expect constraints on some of the inflaton Lagragian 
parameters up to the 0.1\% precision level.  
This is most interesting for building particle physics inspired inflationary 
models.
On the other hand, precision is lost for the determination of
the scalar and tensor tilts on large scales, as well as on the quantity
$C_{10}^T/C_{10}^S$, related by the slow-roll equations to $n_T(k)$.
Finally, in usual inflationary models, the inclusion of polarization 
measurements is known to increase the precision for the parameter extraction. 
In our model, polarization measurements by {\it Planck} are even shown to 
render the extraction of the inflationary parameters up to about 10 times more
accurate.

\section{Acknoledgements}
We thank Alexei A. Starobinsky for illuminating discussions.


\begin{thebibliography}{99}
\bibitem{} Bagla~J.~S., Padmanabhan~T., Narlikar~J.~V., 1996, {\sl 
Crisis in Cosmology --- Observational Constraints on $\Omega_o$ and
$H_o$}, Comments on Astrophysics 18, 275
\bibitem{} Bennett~C.~L. {\it et al.}, 1996, ApJ, 464, L1
\bibitem{} Bond~J.R., Efstathiou~G. and Tegmark~M., 1997, MNRAS, 291, L33
\bibitem{} Bouchet~F.~R., Prunet~S. and Sethi~S.~K., accepted in MNRAS
\bibitem{} Copeland~E.~J., Grivell~I.~J. and Liddle~A.~R., 1998, MNRAS,
in press (astro-ph/9712028)
\bibitem{} Einasto~J., Einasto~M., Gottl\"{o}ber~S., M\"{u}ller~V. 
et al., 1997a, Nature, 385, 139
\bibitem{} Einasto~J., Einasto~M., Frisch~P., Gottl\"{o}ber~S. 
et al., 1997b, MNRAS, 289, 801
\bibitem{} Einasto~J., Einasto~M., Frisch~P., Gottl\"{o}ber~S. 
et al., 1997c, MNRAS, 289, 813
\bibitem{} Eisenstein~D.~J., Hu~W. and Tegmark~M., 1998, submitted to ApJ
(astro-ph/9807130) 
\bibitem{} Gazta\~naga~E. and Baugh~C.~M., 1998, MNRAS 294, 229
\bibitem{} Jungman~G., Kamionkowski~M., Kosowsky~A., Spergel~D.~L.,
1996a, Phys. Rev. Lett., 76, 1007
\bibitem{} Jungman~G., Kamionkowski~M., Kosowsky~A., Spergel~D.~L.,
1996b, Phys. Rev. D, 54, 1332
\bibitem{} Kamionkowski~M., Kosowsky~A. and Stebbins~A., 1997,
Phys.~Rev.~D, 55, 7368
\bibitem{} Kofman~L.~A., Gnedin~N.~Y., Bahcall~N.~A., 1993, ApJ, 413, 1  
\bibitem{} Lesgourgues~J., Polarski~D., Starobinsky~A.~A., 1998a,
MNRAS, 297, 769 (LPS1)
\bibitem{} Lesgourgues~J., Polarski~D., Starobinsky~A.~A., 1998b,
MNRAS, in press (astro-ph/9807019) (LPS2) 
\bibitem{} Lineweaver~C.~H., 1998, ApJ Lett., in press
(astro-ph/9805326)
\bibitem{} Lineweaver~C.~H., Barbosa~D., 1998, ApJ, 496, 624
\bibitem{} Ostriker~J.~P., Steinhardt~P.~J., 1995, Nature, 377, 600 
\bibitem{} Press~W.~H., Flannery~B.~P., Teukolsky~S.~A., Vetterling~W.~T.,
1989, {\it Numerical Recipies}, Cambridge University Press
\bibitem{} Prunet~S., Sethi~S.~K., Bouchet~F.~R., 1998a, to appear in Proc.
Rencontres de Moriond,
{\it Fundamental parameters in Cosmology}
(astro-ph/9803160)
\bibitem{} Prunet~S., Sethi~S.~K., Bouchet~F.~R., 1998b,
in preparation
\bibitem{} Retzlaff~J., Borgani~S., Gottl\"{o}ber~S., Klypin~A.,
M\"{u}ller~V., 1997,
NewA, in press (astro-ph/9709044)
\bibitem{} Seljak~U., Zaldarriaga~M., 1996, ApJ, 469, 7 
\bibitem{} Starobinsky~A.~A., 1992, JETP Lett., 55, 489
\bibitem{} Stompor~R. and Efstathiou~G., 1998, MNRAS, in press
(astro-ph/9805294) 
\bibitem{} Tegmark~M., Taylor~A., Heavens~A., 1997, ApJ, 480, 22
\bibitem{} Wang~Y., Spergel~D.~N., Strauss~M.~A., 1999, ApJ, in press
(astro-ph/9802231)
\bibitem{} White~S.~D.~M., Efstathiou~G., Frenk~C.~S., 
1993, MNRAS, 262, 1023 
\bibitem{} Zaldarriaga~M., Spergel~D.~N., Seljak~U., 1997, ApJ, 488, 1
\end{thebibliography}
\end{document}